# Wavefunctions of Auger Electrons Emitted from Copper Atoms: A Computational Study

Mohamed Assad Abdel-Raouf

Physics *Department, Faculty of Science, UAEU*

*Al-Ain, P.O. Box: 17555, United Arab Emirates..*

## Abstract

An elaborate least-squares variational technique is proposed for obtaining improved continuum wavefunctions of Auger electrons emitted from arbitrary systems the electronic structures of which are determined via one of the computer codes available. A computer program is developed on the basis of this technique for calculating the Auger transition rates using the angular momentum average scheme. The program is employed for investigating the emission of Auger electrons as a consequence of K-shell ionizations of two configurations of copper atoms, (namely $3d^{10}4s^1$ and $3d^9 4s^2$). A comparison is presented between the improved Auger transitions and the ones determined using conventional Hartree-Fock codes. The results emphasize the argument that $(3p \rightarrow 3d3d)$ is the dominant transition in both configurations of the copper atoms. There is a minimal change in the values of the Auger transition rates obtained from these configurations. It is anticipated that the improved continuum wavefunctions provide closer results to the experimental ones.

# 1. Introduction

The creation of holes in different atoms and molecules is one of the most interesting applications of slow positron physics. This process is usually followed by the appearance of a very important phenomenon, namely the emission of highly efficient Auger electrons. Recently, Weiss et al [1] has considered both processes as ultimate spectroscopic tools with wide industrial applications. In particular, these techniques are significant for studying the electronic structure and defects of surfaces of metals and bulks. The main goal of the present work is to develop a novel systematic theoretical scheme which enables the calculation of Auger transition rates from different ionized atoms. Generalization of this scheme to the investigation of metal surfaces should be considered in future. An improvement upon the continuum wavefunctions obtained by conventional Hartree-Fock codes is calculated using a least-squares variational technique [2] in which the Auger electron is affected by the potential of the doubly ionized atom. Previous employment of the same technique [3] for the calculation of the Auger rates emitted by ionized Argon atoms has emphasized the argument that the least-squares results are superior to the Hartree-Fock ones comparative to the experimental data. The second goal of the present work is to implement this scheme in the calculation of Auger transition rates of electrons emitted due to K- shell ionization of the copper atoms. The basic formulae of our approach will be presented in section 2. The results, discussions and conclusions of our investigations are given in section 3.

## 2. Theory

### 2.1. Auger Transition Rates

Inner shell ionization of atoms and molecules could be initiated via electron, positron or photon interactions. In the first case, a relatively high energetic electron collides with an atom (or molecule), kicks out one of its orbital electrons and escape from the interaction region accompanied with the kicked electron. In the second case a positron annihilates with an electron either directly or following a positronium formation. In the third case the inner shell ionization is subjected to photoabsorption process. In all cases a hole is created leading behind to a rearrangement process. The most interesting consequence of this process is the one in which an electron from an outer shell falls to fill the hole. The energy difference between the two states is absorbed by an electron belonging to a less energetic state. Immediately after the absorption the electron will fly out of the interaction region leaving behind a doubly ionized atom. The transition rate, $A^{(0)}_a$, of the flying electron, usually referred to (after its discoverer) as Auger electron, provides us with enough information about the electronic structure of the original target and the history of the process. Theoretically, $A^{(0)}_a$ can be determined by

$$A^{(0)}_a = \frac{2\pi}{\tau} \left| \langle \varphi_3(r_1) \varphi_c(r_2) | \frac{1}{|r_1 - r_2|} | \varphi_1(r_1) \varphi_2(r_2) \rangle \right|^2 , \quad (1)$$

where $\tau = h\, a_0/2\pi e^2 = h^2/4\pi^2 m e^4 = 2.42 \times 10^{-17}$ sec, is the time atomic unit, $\varphi_c$ is the wavefunction of the emitted Auger electron, (to be also referred to as the continuum wavefunction). $\varphi_3$ is the wavefunction describing the electron that filled the hole. The wavefunctions $\varphi_1$ and $\varphi_2$ represent the original states of these two electrons. In the present work a computer code is developed for determining the bound state wavefunctions of the atom using a conventional Hartree-Fock program [4], whilst the wavefunction of the Auger electron is calculated via a least-squares variational approach. The milestone of this code is the evaluation of the interaction potential between the Auger electron and the doubly ionized atom (for more details see [3]).

Following McGuire [5], in the angular momentum average scheme, the Auger rate for the transition $n_1\ell_1 n_2\ell_2 \rightarrow n_3\ell_3 n_c\ell_c$ is connected with the two electrons Auger rate $A^{(0)}_a$ by two different forms:

Case I: $n_1\ell_1 \neq n_2\ell_2$

$$\overline{A_a} = [(h_3+1)(4\ell_1+2-h_1)(4\ell_2+2-h_2)] / [(4\ell_1+2)(4\ell_2+2)]\, A^{(0)}_a(i \rightarrow f), \quad (2)$$

where $h_3$, $h_1$, and $h_2$ are the numbers of holes in the subshells $n_3\ell_3$, $n_1\ell_1$ and $n_2\ell_2$, respectively. Also, the symbols i and $f$ specify an initial and a final state, respectively.

Case II: $n_1\ell_1 = n_2\ell_2$

$$\overline{A_a} = [(h_3+1)(4\ell_1+2-h_1)(4\ell_1+2-h_1)] / [(4\ell_1+2)(4\ell_1+2)]\, A^{(0)}_a(i \rightarrow f), \quad (3)$$

where

$$A^{(0)}_a(n_1\ell_1 n_2\ell_2 \rightarrow n_3\ell_3 n_c\ell_c) = (1/2)(2\ell_c+1)(2\ell_1+1)(2\ell_1+1)\sum_{h,g} hg[\mathrm{Im}(f,g)]^2 \quad (4)$$

h and g are the spin and orbital quantum numbers. The corresponding Auger widths are determined by

$$\Gamma_a(i) = \sum_f \overline{A_a}(i \rightarrow f) \quad (5)$$

## 2.2 Improved Continuum Wavefunctions

The exact continuum electron wavefunction $|\phi_c\rangle$ has to satisfy Schrödinger's equation

$$(H-E)|\phi_c\rangle = |0\rangle, \qquad (6)$$

where H and E are the total Hamiltonian and energy, respectively. The total Hamiltonian H is given by

$$H = \frac{-h^2}{2m}\nabla_r^2 + V(r), \qquad (7)$$

where **V(r)** is the potential seen by the continuum electron at a distance r from the infinitely heavy nucleus of the considered atom. Thus, **V(r)** has the form

$$V(r) = \frac{-Z_{eff}\, e^2}{r} + V_c^q(r), \qquad (8)$$

where $Z_{eff} = Z - N$ (Z is the nuclear charge of an atom q and N is the number of remaining electrons), $V_c^q(r)$ is a screening core potential defined by

$$V_c^q(r) = V_{c\ Coul}^q(r) + V_{c\ cex}^q(r), \qquad (9)$$

$V_{c\ Coul}^q(r)$ and $V_{c\ cex}^q(r)$ are the Coulomb and exchange parts of the total potential. According to Abdel-Raouf[11] (1988), these potentials can be derived and having the following forms

$$V^q_{c\,coul}(r) = \sum_{j=1}^{M^q} {}' N^q_j \langle \Phi^q_j(r_i) \left| \frac{2}{|r-r_i|} - \frac{2}{r} \right| \Phi^q_j(r_i) \rangle, \tag{10}$$

where $M^q$ is the number of core orbitals of an atom or ion q and $N^q_j$ is the number of electrons occupying the orbital j and each electron is at a distance $r_i$ from the nucleus and $\Phi^q_j(r_i)$ is the wavefunction of the i th electron in the orbital j. The prime on the sum sign means that the term $\frac{-2}{r}$ is repeated for each j. The exchange part of the core potential is defined by

$$V^q_{cex}(r) = -\sum_{j=1}^{M^q} \langle \Phi^q_c(r_i) \left| \frac{2}{|r-r_i|} \right| \Phi^q_j(r_i) \rangle, \tag{11}$$

where c is employed for distinguishing the wavefunction of Auger electron (continuum electron). $|\phi_c\rangle$, in eq. (5), can be expanded by

$$|\phi_c\rangle = |S\rangle + K_{11}|C\rangle + |\phi\rangle, \tag{12}$$

where $K_{11}$ is the tangent of the s-wave phase shift $\eta_0$, $|S\rangle$ is the regular part, $|C\rangle$ is the irregular part with an associated cut-off function for excluding the singularity at the origin and $|\phi\rangle$ is a Hilbert-space wavefunction describing possible virtual states composed of the continuum electron and the ion. It is the superposition of bound state functions $|\chi_i\rangle$ which go asymptotically to zero. Thus we have

$$|\phi\rangle = \sum_{i=1}^{n} d_i |\chi_i\rangle \tag{13}$$

The least-squares variational method proceeds (Abdel-Raouf[12], 1984) by defining a trail wavefunction such that

$$|\phi_c^{(n)}\rangle = |S\rangle + K_{11}|C\rangle + \sum_{i=1}^{n} d_i |\chi_i\rangle \qquad (14)$$

The wavefunctions $|S\rangle$, $|C\rangle$ and $|\chi_i\rangle$ of eq. (14) are choosen as

$$S = \frac{\sin(Kr)}{\sqrt{K}}, \quad C = \frac{\cos(Kr)(1-e^{-\alpha r})}{\sqrt{K}} \quad \text{and} \quad \chi_i = r^i e^{-\alpha r}.$$

The variational parameters $K_{11}$ and $d_i$ (i=1, 2, ..., n) are determined by considering the following projections

$$\langle S | H\text{-}E | \phi_c^n \rangle = V_1,$$

$$\langle C | H\text{-}E | \phi_c^n \rangle = V_2 \quad \text{and} \qquad (15)$$

$$\langle \chi_i | H\text{-}E | \phi_c^n \rangle = V_{i+2}, \qquad i = 1, 2, ..., n.$$

The $V_i$'s are subjected to the following variational principle

$$\delta \sum_{i=1}^{n+2} V_i^2 = 0, \qquad (16)$$

and the variational parameters are obtained by applying this variational principle. Thus, the final form of $|\phi_c^{(n)}\rangle$ can be written as

$$|\phi_c^{(n)}\rangle = \frac{1}{\Delta:n}\left\{ \left|\begin{array}{c}|S\rangle \\ M_\cdot^s\end{array}\right| \frac{|\chi_1\rangle \cdots |\chi_n\rangle}{\Delta:n} + K_{11}^n \left|\begin{array}{c}|C\rangle \\ M_\cdot^c\end{array}\right| \frac{|\chi_1\rangle \cdots |\chi_n\rangle}{\Delta:n} \right\} \qquad (17)$$

where $\quad M_:^S = \begin{bmatrix} (S: \chi_i) \\ (S: \chi_2) \\ . \\ . \\ . \\ . \\ (S: \chi_n) \end{bmatrix}$, $\quad M_:^C = \begin{bmatrix} (C: \chi_1) \\ (C: \chi_2) \\ . \\ . \\ . \\ . \\ (C: \chi_n) \end{bmatrix}$

and $\quad \Delta: n = \begin{bmatrix} (\chi_1: \chi_1) & (\chi_1: \chi_2)\cdots\cdots(\chi_1: \chi_n) \\ (\chi_2: \chi_1) & (\chi_2: \chi_2)\cdots\cdots(\chi_2: \chi_n) \\ . \\ . \\ . \\ (\chi_n: \chi_1) & (\chi_n: \chi_2)\cdots\cdots(\chi_n|\chi_n) \end{bmatrix}$ (17)

The matrix element (f: g) is defined by

$$(\mathbf{f: g}) = \sum_{k=-1}^{n} (f|\chi_k)(\chi_k|g).$$

Particularly, we have $|\chi_{-1}\rangle = |S\rangle$ and $|\chi_0\rangle = |C\rangle$, the $\chi_i$ s, $i \geq 1$, are the Hilbert-space wavefunctions defined above and e.g. $(f|\chi_i) = \langle f|H - E|\chi_j\rangle$.

Remark that the variational process can be carried out by assuming a certain form of $|\Phi_c>$ appearing at eq. (11). It is suggested here to employ the continuum wavefunction used by the distorted wave approximation as original Hartree Fock. This provides us by the first order iteration of $|\Phi_c>$, eq. (17), and consequently the first order iteration of Auger transition rate, eq. (4). An iterative second order improvement upon the last $|\Phi_c>$ can be achieved by introducing it into eq. (11), in order to obtain new set of values of the potentials V(r) at all points of the space, and apply the above mentioned least-square variational technique. The new $|\Phi_c>$ leads to a second order iteration of the transition rate, eq. (4). According to the available computer facilities, we can calculate automatically higher order iterations.

## 3. Results and Discussions

Auger transition rates $A_a$, Auger widths $\Gamma_a$, and energies of continuum Auger electrons $E_c$ corresponding to the initial K-vacancies are calculated for the two forms of copper atoms Cu (I) and Cu (II) using the angular momentum average ( AMA ) scheme. The probable transitions are included. Detailed data for Auger transition rates $A_a$ (measured in sec-l) and the corresponding continuum electron energies $E_c$ (measured in Ry) are given in Table 1 for Cu (I) for K-Auger transitions. Similar data are presented in Table 2 for Cu (II) case.

From the Tables we conclude the following points:

(1) Comparison between columns 4 and 5 in each Table supports the argument mentioned in the introduction (see also [3]) that valuable improvements upon the Auger transition rates calculated from Hartree – Fock program can be achieved using the least-squares variational method for improving the continuum wavefunctions of the Auger electrons.

(2) The Auger transition $A_a$ (3p $\to$ 3d3d) is the dominant one in both configurations. It occurs at continuum electron energy 67.6 eV for Cu (I), and $E_c$ = 60.39 eV for Cu (II).

(3) There is a small difference in the results of the Auger rates $A_a$ between both electronic configuration of the copper atom Cu (I) and Cu (II).

Table 1: Comparison between Auger transition rates calculated directly from the direct Hartree - Fock program (DHF) and the Improved ones (column 5) in sec$^{-1}$ for initial states of Cu (I), the energy of the emitted electron (in Ry) and the angular momentum of this electron ($\ell_c$).

| Cu (I) Initial State : $1s^1 2s^2 2p^6 3s^2 3p^6 3d^{10} 4s^1$ | | | | |
|---|---|---|---|---|
| Final State | $\ell_c$ | $E_c$ (Ry) | $\overline{A_a}$ (Sec$^{-1}$) DHF | $\overline{A_a}$ (Sec$^{-1}$) Improved HF |
| **$1s^2 2p^6 3s^2 3p^6 3d^{10} 4s^1$** | 0 | 441.3 | 4.3299x10$^{13}$ | 4.3304 x10$^{13}$ |
| $1s^2 2s^2 2p^4 3s^2 3p^6 3d^{10} 4s^1$ | 0 | 468.08 | 2.33219 x10$^{13}$ | 2.3328 x10$^{13}$ |
| $1s^2 2s^2 2p^6 3s^2 3p^6 3d^{10} 4s^1$ | 0 | 593.38 | 1.2530 x10$^{12}$ | 1.2535 x10$^{12}$ |
| $1s^2 2s^2 2p^6 3s^2 3p^6 3d^{10} 4s^1$ | 0 | 600.67 | 8.0733 x10$^{11}$ | 8.0739 x10$^{11}$ |
| $1s^1 2s^2 2p^6 3s^2 3p^4 3d^8 4s^1$ | 0 | 656.91 | 5.2068 x10$^7$ | 5.2073 x10$^7$ |
| | | | | |
| $1s^2 2s^1 2p^5 3s^2 3p^6 3d^{10} 4s^1$ | 1 | 451.66 | 1.7691 x10$^{13}$ | 1.7691 x10$^{13}$ |
| $1s^2 2s^1 2p^6 3s^1 3p^6 3d^{10} 4s^1$ | 0 | 516.39 | 8.1176 x10$^{12}$ | 8.1188 x10$^{12}$ |
| $1s^2 2s^1 2p^6 3s^2 3p^5 3d^{10} 4s^1$ | 1 | 520.97 | 1.821 x10$^{13}$ | 1.822 x10$^{13}$ |
| $1s^2 2s^1 2p^6 3s^2 3p^6 3d^9 4s^1$ | 2 | 571.97 | 1.1006 x10$^{12}$ | 1.1010 x10$^{12}$ |
| $1s^2 2s^1 2p^6 3s^2 3p^6 3d^9$ | 0 | 548.35 | 2.1935 x10$^{11}$ | 2.1937 x10$^{11}$ |
| $1s^2 2s^2 2p^5 3s^1 3p^6 3d^{10} 4s^1$ | 1 | 526.72 | 1.5861 x10$^{13}$ | 1.5879 x10$^{13}$ |
| $1s^2 2s^2 2p^5 3s^2 3p^5 3d^{10} 4s^1$ | 0 | 530.40 | 1.3412 x10$^{13}$ | 1.3415 x10$^{13}$ |
| $1s^2 2s^2 2p^5 3s^2 3p^6 3d^9 4s^1$ | 1 | 558.51 | 1.767 x10$^{11}$ | 1.769 x10$^{11}$ |
| | | | | |
| $1s^2 2s^2 2p^6 3s^2 3p^6 3d^{10}$ | 1 | 558.69 | 1.9985 x10$^{11}$ | 1.9987 x10$^{11}$ |
| $1s^2 s^2 2p^6 3s^2 3p^6 3d^{10} 4s^1$ | 1 | 597.01 | 3.66659(x10$^{10}$ | 3.66739 x10$^{10}$ |
| $1s^2 2s^2 2p^6 3s^1 3p^6 3d^9 4s^1$ | 2 | 625.15 | 1.170 x10$^{11}$ | 1.195 x10$^{11}$ |
| $1s^2 2s^2 2p^6 3s^1 3p^6 3d^{10}$ | 0 | 624.45 | 4.6045x10$^{10}$ | 4.6048 x10$^{10}$ |
| $1s^2 2s^2 2p^6 3s^2 3p^5 3d^{10}$ | 1 | 628.10 | 5.1105 x10$^{10}$ | 5.1401 x10$^{10}$ |
| $1s^2 2s^2 2p^6 3s^2 3p^6 3d^9$ | 2 | 656.2 | 1.3289 x10$^9$ | 1.3485 x10$^9$ |
| | | | $\Gamma_a$=7.445 x10$^{14}$ | $\Gamma_a$=7.547 x10$^{14}$ |

Table2: Comparison between Auger transition rates calculated directly from the direct Hartree - Fock program (DHF) and the Improved ones (column 5) in sec$^{-1}$ for initial states of Cu (II) , the energy of the emitted electron (in Ry) and the angular momentum of this electron ($\ell_c$).

| Cu (II)  Initial State : $1s^1 2s^2 2p^6 3s^2 3p^6 3d^9 4s^2$ | | | | |
|---|---|---|---|---|
| Final State | $\ell_c$ | $E_c$ (Ry) | $\overline{A_a}$ (Sec$^{-1}$) DHF | $\overline{A_a}$ (Sec$^{-1}$) Improved HF |
| $1s^2 2p^6 3s^2 3p^6 3d^9 4s^2$ | 0 | 441.18 | 9.6605x10$^{13}$ | 9.6635x10$^{13}$ |
| $1s^2 2s^2 2p^4 3s^2 3p^6 3d^9 4s^2$ | 0 | 461.96 | 2.33587x10$^{13}$ | 2.33595 x10$^{13}$ |
| $1s^2 2s^2 2p^6 3p^6 3d^9 4s^2$ | 0 | 593.17 | 2.6648 x10$^{12}$ | 2.6666 x10$^{12}$ |
| $1s^2 2s^2 2p^6 3s^2 3p^4 3d^9 4s^2$ | 0 | 600.44 | 5.9835 x10$^{11}$ | 5.9836 x10$^{11}$ |
| $1s^2 2s^2 2p^6 3s^2 3p^6 3d^7 4s^2$ | 0 | 655.5 | 2.8959 x10 | 2.8964 x10$^7$ |
| $1s^2 2s^2 2p^6 3s^2 3p^6 3d^9 4s^2$ | 1 | 505.34 | 9.4169x10$^{13}$ | 9.4174 x10$^{13}$ |
| $1s^2 2s^1 2p^6 3s^1 3p^6 3d^9 4s^2$ | 0 | 565.32 | 5.5562 x10$^{12}$ | 5.5580 x10$^{12}$ |
| $1s^2 2s^1 2p^6 3s^2 3p^5 3d^9 4s^2$ | 1 | 520.80 | 1.9904x10$^{13}$ | 1.9929 x10$^{13}$ |
| $1s^2 2s^1 2p^6 3s^2 3p^6 3d^8 4s^2$ | 2 | 548.35 | 6.0191 x10$^{12}$ | 6.0251 x10$^{12}$ |
| $1s^2 2s^1 2p^6 3s^2 3p^6 3d^8 4s^2$ | 0 | 548.92 | 1.3819 x10$^{11}$ | 1.3877 x10$^{11}$ |
| $1s^2 2s^2 2p^5 3s^1 3p^6 3d^9 4s^2$ | 1 | 527.50 | 1.5831x10$^{13}$ | 1.5873 x10$^{13}$ |
| $1s^2 2s^2 2p^5 3s^2 3p^5 3d^9 4s^2$ | 0 | 531.18 | 1.3312x10$^{13}$ | 1.3344 x10$^{13}$ |
|  | 2 |  | 1.4113(14) | 1.4150 (14) |
| $1s^2 2s^2 2p^5 3s^2 3p^6 3d^8 4s^2$ | 1 | 558.69 | 1.6522 x10$^{11}$ | 1.6562 x10$^{11}$ |
| $1s^2 2s^2 2p^5 3s^2 3p^6 3d^9 4s^1$ | 1 | 559.79 | 2.9211 x10$^{11}$ | 2.9411 x10$^{11}$ |
| $1s^2 2s^2 2p^6 3s^1 3p^5 3d^9 4s^2$ | 1 | 596.79 | 2.747 x10$^{12}$ | 2.755 x10$^{12}$ |
| $1s^2 2s^2 2p^6 3s^2 3p^6 3d^9 4s^2$ | 2 | 624.33 | 7.7606 x10$^{10}$ | 7.7630 x10$^{10}$ |
| $1s^2 2s^2 2p^6 3s^1 3p^6 3d^8 4s^1$ | 0 | 624.97 | 2.3182 x10$^{10}$ | 2.3252 x10$^{10}$ |
| $1s^2 2s^2 2p^6 3s^2 3p^5 3d^8 4s^2$ | 1 | 627.97 | 1.8731 x10$^{10}$ | 1.8791 x10$^{10}$ |
| $1s^2 2s^2 2p^6 3s^1 3p^5 3d^8 4s^1$ | 1 | 628.65 | 5.0597 x10$^{10}$ | 5.0643 x10$^{10}$ |
| $1s^2 2s^2 2p^6 3s^2 3p^6 3d^8 4s^1$ | 2 | 656.2 | 1.337 x10$^9$ | 1.704 x10$^9$ |
| $1s^2 2s^2 2p^6 3s^2 3p^6 3d^9$ | 0 | 656.8 | 7.828 x10$^8$ | 7.905 x10$^8$ |
|  |  |  | $\Gamma_a$=7.0988x10$^{14}$ | $\Gamma_a$=7.119(14) |